\documentclass[a4paper, oneside, twocolumn, notitlepage, 10pt]{extarticle_ecoc}
\usepackage{ecoc}
\usepackage{amsmath,amssymb}
\usepackage{pgfplotstable}
\usepackage{subcaption}
\usepackage{siunitx}
\usepackage{url}
\usetikzlibrary{decorations.text}
\definecolor{C1}{RGB}{031, 119, 180} 
\definecolor{C2}{RGB}{255, 127, 014} 
\definecolor{C3}{RGB}{044, 160, 044} 
\definecolor{C4}{RGB}{215, 039, 040} 
\definecolor{C6}{RGB}{148, 103, 189} 
\definecolor{C100}{RGB}{140, 086, 075} 
\definecolor{C7}{RGB}{227, 119, 194} 
\definecolor{C8}{RGB}{127, 127, 127} 
\definecolor{C9}{RGB}{188, 189, 034} 
\definecolor{C5}{RGB}{023, 190, 207} 

\definecolor{C11}{RGB}{174, 199, 232} 
\definecolor{C12}{RGB}{255, 187, 120} 
\definecolor{C13}{RGB}{152, 223, 138} 
\definecolor{C14}{RGB}{255, 152, 150} 
\definecolor{C15}{RGB}{197, 176, 213} 
\definecolor{C16}{RGB}{196, 156, 148} 
\definecolor{C17}{RGB}{247, 182, 210} 
\definecolor{C18}{RGB}{199, 199, 199} 
\definecolor{C19}{RGB}{219, 219, 141} 
\definecolor{C20}{RGB}{158, 218, 229} 
\pgfplotscreateplotcyclelist{foo}{
        {C1,mark=*},
        {C2,mark=square*},
        {C3, mark = triangle*},
        {C4,mark=+},
        {C5, mark = diamond*},
        {C6, mark = x}
      }

\begin{document}
\selectlanguage{english}    


\title{Short Blocklength Error Correction Codes for Continuous-Variable Quantum Key Distribution \vspace{-3mm}}%


\author{
    Kadir G\"{u}m\"{u}\c{s}\textsuperscript{(1)}, João dos Reis Frazão\textsuperscript{(1)},
    Boris \v{S}kori\'{c} \textsuperscript{(2)}, \\
    Gabriele Liga \textsuperscript{(3)},
    Aaron Albores-Mejia \textsuperscript{(1)}, 
    Thomas Bradley \textsuperscript{(1)}, and
    Chigo Okonkwo\textsuperscript{(1)} \vspace{-2mm}
}

\maketitle                  


\begin{strip}
    \begin{author_descr}

        \textsuperscript{(1)} High-Capacity Optical Transmission Laboratory, Electro-Optical Communications Group, Eindhoven
University of Technology, The Netherlands, 
        \textcolor{blue}{\uline{k.gumus@tue.nl}} 

        \textsuperscript{(2)} Dept. of Mathematics and Computer Science,
Eindhoven University of Technology, The Netherlands

        \textsuperscript{(3)} Information and Communication Theory Lab,
Eindhoven University of Technology, The Netherlands
\vspace{-3mm}
    \end{author_descr}
\end{strip}

\renewcommand\footnotemark{}
\renewcommand\footnoterule{}


\begin{strip}
    \begin{ecoc_abstract}
    We introduce a two-step error correction scheme for reconciliation in continuous-variable quantum key distribution systems. Using this scheme, it is possible to use error correction codes with small blocklengths (1000 bits), increasing secret key rates at a distance of 140km by up to 7.3 times.
\textcopyright2024 The Author(s)
\vspace{-3mm}
    \end{ecoc_abstract}
\end{strip}

\section{Introduction}
Quantum key distribution (QKD) is a method for securely sharing keys between two parties, Alice and Bob, in the presence of an eavesdropper with access to a quantum computer \cite{Bennett_2014}. Continuous-variable (CV) QKD is a variant of QKD allowing for sharing secret keys using conventional telecommunication equipment \cite{GG02}. Compared to discrete-variable QKD, two disadvantages that CV-QKD has are that the transmission distance is limited and that the implementation of the reconciliation is significantly more difficult \cite{yang2023information}.

Reconciliation is part of the post-processing for a QKD system and is where the keys are generated in the case of CV-QKD. Works focusing on reconciliation \cite{Milicevic_2018, gumucs2025rate, cil2024rate, mani2021multiedge} use long blocklengths ($10^5 - 10^6$ bits), as it increases performance. These codes, however, are quite complex to implement on application-specific hardware. Additionally, for higher reconciliation efficiencies $\beta \geq 98\%$, the frame error rate (FER) approaches $1$ very quickly, reducing key rates at longer distances.

A potential solution to these problems is using short blocklength error correction codes ($10^3 - 10^4$ bits). These codes require fewer decoding iterations and take up fewer resources on hardware. In general, at high $\beta$, the FER is better than codes with long blocklengths. However, the error detection after decoding, which is often chosen to be a cyclic redundancy check (CRC) \cite{Milicevic_2018}, significantly affects $\beta$ and makes short blocklength error correction codes infeasible.

This paper proposes a protocol for using short blocklength error correction codes for CV-QKD.
We show that with this method, it is possible to perform reconciliation at high $\beta$ ($\geq 99\%$) with reasonable FERs ($\leq 87\%$). Increases in secret key rates over a distance of 140km of up to 7.3 times are shown.

\begin{figure*}[t!]
    \centering
    \resizebox{\linewidth}{!}{\tikzstyle{gradientc1} = [bottom color=C1!5, top color=C1!15]
\tikzstyle{gradientc4} = [bottom color=C4!5, top color=C4!15]    

\begin{tikzpicture}
\draw[dashed, very thick, C8](-10.5,-4.6) rectangle (14.5,-6.4);
\node[C8] at (2,-5.5){\Large Classical Channel};
    \draw[dashed,very thick, C1](-10.5,-4.5) rectangle (16.5,-1);

    \node[C1] at (-9.2,-1.7){\Huge Alice};

    \draw[rounded corners,thick,C1,gradientc1] (10,-2) rectangle node[midway, align = center]{Multi-dimensional\\ Reconciliation \\ Demapper} (14,-4);

    \draw[rounded corners,thick,C1,gradientc1] (4,-2) rectangle node[midway, align = center]{Short Blocklength\\ Low Rate \\ Decoder} (8,-4);

    \draw[rounded corners,thick,C1,gradientc1] (-8,-2) rectangle node[midway, align = center]{High Rate\\ Long Blocklength \\ Decoder} (-4,-4);
    \draw[dashed,very thick,C4](-10.5,-6.5) rectangle (16.5,-10);

    \node[C4] at (-9.2,-9.3){\Huge Bob};

    \draw[rounded corners,thick,C4,gradientc4] (4,-9) rectangle node[midway, align = center]{Short Blocklength\\ Low Rate \\ Encoder} (8,-7);

    \draw[rounded corners,thick,C4,gradientc4] (-2,-7) rectangle node[midway, align = center]{Quantum Random \\ Number Generator} (2,-9);
    
    \draw[rounded corners,thick,C4,gradientc4] (10,-7) rectangle node[midway, align = center]{Multi-dimensional\\ Reconciliation \\ Mapper} (14,-9);
        
    \draw[rounded corners,thick,C4,gradientc4] (-4,-7) rectangle node[midway, align = center]{High Rate\\ Long Blocklength \\ Encoder} (-8,-9);


\draw[thick, <->,C6](15,-4) -- node[midway,right]{$I_{AB}$}(15,-7);

\draw[thick,<-,C1](14,-2.5) -- node[midway,above]{$\mathbf{x}_1$} (16,-2.5);

\node[rotate = 90,C1] at (15,-3){$\cdots$};

\draw[thick,<-,C1](14,-3.5) -- node[midway,below]{$\mathbf{x}_K$} (16,-3.5);

\draw[thick,<-,C4](14,-7.5) -- node[midway,above]{$\mathbf{y}_1$} (16,-7.5);

\node[rotate = 90,C4] at (15,-8){$\cdots$};

\draw[thick,<-,C4](14,-8.5) -- node[midway,below]{$\mathbf{y}_K$} (16,-8.5);

\draw[thick,->,C8](11,-7) -- node[midway,left]{$\mathbf{m}_1$} (11,-4);

\node[C8] at (12,-5.5){$\cdots$};

\draw[thick,->,C8](13,-7) -- node[midway,right]{$\mathbf{m}_K$} (13,-4);

\draw[thick,->,C4](2,-7.5) -- node[midway,above]{$\mathbf{s}_1$} (4,-7.5);

\node[rotate = 90, C4] at (3,-8){$\cdots$};

\draw[thick,->,C4](2,-8.5) -- node[midway,below]{$\mathbf{s}_K$} (4,-8.5);

\draw[thick,->,C4](-2,-7.5) -- node[midway,above]{$\mathbf{s}_1$} (-4,-7.5);

\node[rotate = 90, C4] at (-3,-8){$\cdots$};

\draw[thick,->,C4](-2,-8.5) -- node[midway,below]{$\mathbf{s}_K$} (-4,-8.5);

\draw[thick,->,C4](8,-8.5) -- node[midway,below]{$\mathbf{c}_K$} (10,-8.5);

\node[rotate = 90,C4] at (9,-8){$\cdots$};

\draw[thick,->,C4](8,-7.5) -- node[midway,above]{$\mathbf{c}_1$} (10,-7.5);

\draw[thick,<-,C1](8,-2.5) -- node[midway,above]{$\mathbf{l}_1$} (10,-2.5);

\node[rotate = 90,C1] at (9,-3){$\cdots$};

\draw[thick,<-,C1](8,-3.5) -- node[midway,below]{$\mathbf{l}_K$} (10,-3.5);

\draw[thick,->,C1](4,-2.5) -- node[midway,above]{$\mathbf{\hat{s}}_{idx_1}$} (-4,-2.5);

\node[rotate = 90,C1] at (0,-3){$\cdots$};

\draw[thick,->,C1](4,-3.5) -- node[midway,below]{$\mathbf{\hat{s}}_{idx_{A}}$} (-4,-3.5);

\draw[thick,<-,C8](-6,-7) -- node[midway,left]{$idx_1$} (-6,-4);

\node[C8] at (-5.375,-5.5){$\cdots$};

\draw[thick,<-,C8](-4.75,-7) -- node[midway,right]{$idx_{A}$} (-4.75,-4);

\draw[thick,<-,C8](-7.25,-4) -- node[midway,left]{$\mathbf{p}$} (-7.25,-7);

\draw[thick,->,C1] (-8,-3) -- node[midway,above]{$\hat{\mathbf{w}}$}(-10,-3);

\draw[thick,->,C4] (-8,-8) -- node[midway,above]{$\mathbf{w}$}(-10,-8);
\end{tikzpicture}}
    \caption{An overview of short-blocklength reconciliation.}
    \label{fig:SBR}
\end{figure*}
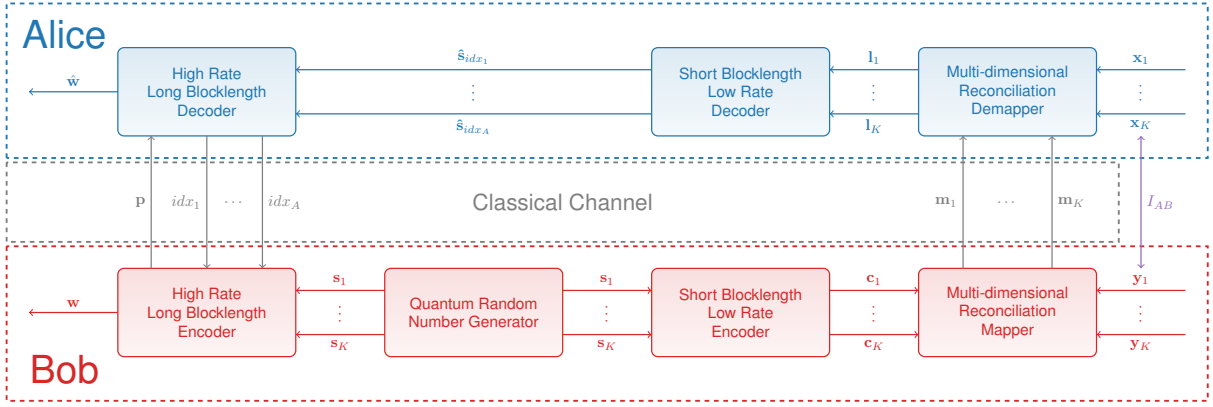

\vspace{-2mm}
\section{Multi-dimensional Reconciliation}
Multi-dimensional reconciliation is a commonly used reconciliation protocol for CV-QKD \cite{Leverrier_2008}. Our protocol, as shown in Fig. \ref{fig:SBR}, is based on multi-dimensional reconciliation but includes a second decoding step. We first describe conventional multi-dimensional reconciliation.

At the start of reconciliation Alice has a string $\mathbf{x}$ of transmitted quadrature values, $\mathbf{x} = [x_1, x_2, \cdots, x_N]$, 
where $N$ is the blocklength of the code, 
and $\mathbb{E}[x_i^2] = \frac{1}{2}$. 
Bob has a string of measured quadrature values $\mathbf{y} = [y_1, y_2, \cdots y_N]$, which is a noisy version of $
\mathbf{x}$. The quantum channel over which Alice transmits the states is assumed to be an additive white Gaussian noise channel; therefore, $y_i = x_i + z_i$. where $z_i \sim \mathcal{N}(0,\sigma_z^2/2)$, where $\sigma_z^2$ is the channel noise variance over both quadratures of the quantum state. 

Using a random number generator, Bob generates a string of bits $\mathbf{s} = [s_1, s_2, \cdots, s_k]$, where $k = R\cdot N$ with $R$ being the rate of the code. 
Using an encoder, Bob calculates the codeword $\mathbf{c}$ belonging to $\mathbf{s}$. 
The codeword is then converted to a sequence of BPSK symbols $\mathbf{u}$ through $u_i = (-1)^{c_i}$. 
Bob generates $\mathbf{m}\in{\mathbb R}^N$ using a mapping function $\mathbf{m} = M(\mathbf{u},\mathbf{y})$ and transmits $\mathbf{m}$ over the classical channel to Alice. 
Alice recovers $\mathbf{m}$, and applies the inverse mapping function to obtain $\mathbf{r} = M^{-1}(\mathbf{m},\mathbf{x})$, which is a noisy version of $\mathbf{u}$. 

Based on $\mathbf{r}$, Alice calculates the log-likelihood ratios $\mathbf{llr}$ 
and uses them as input to her decoder. At the output of her decoder, she obtains a codeword estimate $\hat{\mathbf{c}}$. She then confirms whether $\hat{\mathbf{c}}$ is a valid codeword. 
This confirmation step is code-dependent, and in the case of LDPC codes, this would be equivalent to checking whether the syndrome is 0, i.e., $\hat{\mathbf{c}} \mathbf{H}^T = \mathbf{0}$, where $\mathbf{H}$ is the parity check matrix of the code. 
If the codeword is not valid, a frame error has occurred, and the frame is discarded. Otherwise, $\hat{\mathbf{c}}$ is a valid codeword, but this does not guarantee that $\hat{\mathbf{c}} = \mathbf{c}$.

In conventional reconciliation, an error detection step takes place after error correction. The most common form of error detection is using a cyclic redundancy check (CRC) \cite{Milicevic_2018}. The last $n_{crc}$ bits of the estimated information bits $\hat{\mathbf{s}}$ are transmitted by Alice over the classical channel to Bob. Bob compares these bits to the last $n_{crc}$ information bits $\mathbf{s}$ on his side. If the bits do not match, reconciliation fails, and the frame is discarded. If they are the same, Bob confirms this information with Alice and the information bits of the frame $
\mathbf{s}$ and $\hat{\mathbf{s}}$ are kept for privacy amplification. Importantly, the revealed bits are discarded when accepting the frame, as these are known to Eve.  This has a slight effect on the reconciliation efficiency, as $n_{crc}$ information bits are discarded:
\begin{equation}
    \beta = \frac{R\cdot N - n_{crc}}{N I_{AB}},
\end{equation}
where $I_{AB}$ is the mutual information between $\mathbf{x}$ and $\mathbf{y}$. 

\begin{figure}[b!]
    \centering
      \begin{tikzpicture}[>=latex]
\begin{axis}[
every axis/.append style={font=\small},
tick label style={font=\footnotesize},
xlabel = Number of CRC bits $n_{crc}$ (bits),
ylabel = Reduction in $\beta$ (\%),
xmin = 1, xmax =32,
ymin = 0, ymax = 100,
xmode = log,
log basis x=2,
x tick label style={yshift= -1mm},
y tick label style={xshift= -1mm},
ylabel shift = 1mm,
xlabel shift = 1mm,
set layers, mark layer=axis tick labels,
grid = major,
width = \linewidth,
 height = 6cm,
 xlabel near ticks,  
 ylabel near ticks,  
 xticklabel style={/pgf/number format/fixed},
every axis plot/.append style={thick},legend style={at={(0.1,0.7)},anchor=west, font = \scriptsize,row sep=-0.75ex,inner sep=0.2ex,fill opacity = 0.6, text opacity = 1},
legend cell align={left},
cycle list name = foo
]

\pgfplotstableread{Figures/Penalty_CRCs.txt}
\datatable
\pgfplotsinvokeforeach {1,2,3,4,5,6}{
\addplot+
         table
         [
          x expr=\thisrowno{0}, 
          y expr=\thisrowno{#1} 
         ] {\datatable};
}
\addlegendentry{$N = 10^3$}
\addlegendentry{$N = 2\cdot10^3$}
\addlegendentry{$N = 5\cdot 10^3$}
\addlegendentry{$N = 10^4$}
\addlegendentry{$N = 10^5$}
\addlegendentry{$N = 10^6$}

\end{axis}
\end{tikzpicture}
    \caption{The reduction in $\beta$ vs. $n_{crc}$ for different error correction code blocklengths $N$.}
    \label{fig:CRC}
\end{figure}
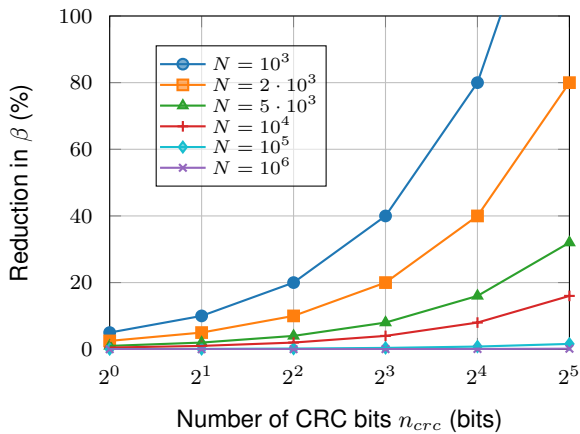

The SKR for multi-dimensional reconciliation can then be given by:
\begin{equation}
    \text{SKR} = (1-\text{FER})(\beta I_{AB} - \chi_{BE} - \Delta_n), 
\end{equation}
where  $\chi_{BE}$ is the Holevo information, and $\Delta_n$ is the finite-size penalty which is dependent on the privacy amplification blocklength $N_{privacy}$ \cite{gumucs2025rate}.

In Fig. \ref{fig:CRC} we show how $n_{crc}$ influences $\beta$  depending on $N$ for an $R = \frac{1}{50}$ type-based protograph (TBP) LDPC code \cite{gumucs2021low}. As $N$ decreases, the effect on $\beta_{crc}$ increases. Even for $n_{crc} = 1$, a reduction in $\beta$ of 5\% occurs when $N = 1000$. Hence, the use of small blocklength error correction codes becomes infeasible. Therefore, we have developed a reconciliation protocol where short blocklength error correction codes can be used, without applying a heavy penalty on $\beta$.

\vspace{-1mm}
\section{Short-Blocklength Reconciliation}
Instead of a CRC, we use a second hard-decision high-rate error correction code, which effectively operates as an outer code, to remove any residual errors between $\hat{\mathbf{s}}$ and $\mathbf{s}$. 
Let the blocklength of the outer code be $N_{out}$, which is a multiple of $k$. We do multi-dimensional reconciliation until $A = N_{out}/k$ frames have been accepted.
 
The number of reconciliation attempts $K$ for this to happen is on average equal to $\mathbb{E}[K] = \frac{N_{out}}{k\cdot (1-\text{FER})}$. These bits, $\mathbf{w} = [\mathbf{s}_{idx_1},\mathbf{s}_{idx_2}, \cdots, \mathbf{s}_{idx_{A}}]$, where $\mathbf{idx}$ denotes the indices of the accepted frames, are not a codeword of the outer code, hence the syndrome is not equal to $\mathbf{0}$. 

Alice sends $\mathbf{idx}$ over the classical channel to Bob. 
Bob computes the syndrome $\mathbf{p}$ of $\mathbf{w}$ and sends the syndrome under one-time pad encryption to Alice.
Alice attempts to recreate $\mathbf{w}$ using the information bits of the accepted frames $\hat{\mathbf{w}} = [\hat{\mathbf{s}}_{idx_1},\hat{\mathbf{s}}_{idx_2}, \cdots, \hat{\mathbf{s}}_{idx_{N_{out}/N}}]$. As there might be bit flips between $\mathbf{w}$ and $\hat{\mathbf{w}}$, Alice uses $\mathbf{p}$ and the decoder of the outer code to remove residual bit errors. Afterwards, Alice and Bob are left with  $\hat{\mathbf{w}}$ and $\mathbf{w}$, which are the same with high probability, and they will use these bits for privacy amplification. 

 Having to one-time pad $\mathbf{p}$ requires some key material, which has to be deducted from the overall SKR. We can write this as a penalty  $R_{out}$ on $\beta$:
\begin{equation}
    \beta = \frac{R\cdot R_{out}}{I_{AB}}.
\end{equation}
From our simulations, for all $\beta$ and $N$, we observed that the bit error rate (BER) between $\mathbf{w}$ and $\hat{\mathbf{w}}$ is relatively small (BER << $10^{-4}$). Hence, the rate of the outer code $R_{out}$ can be very high, $R_{out} > 0.999$, and the penalty to the reconciliation efficiency is negligible.

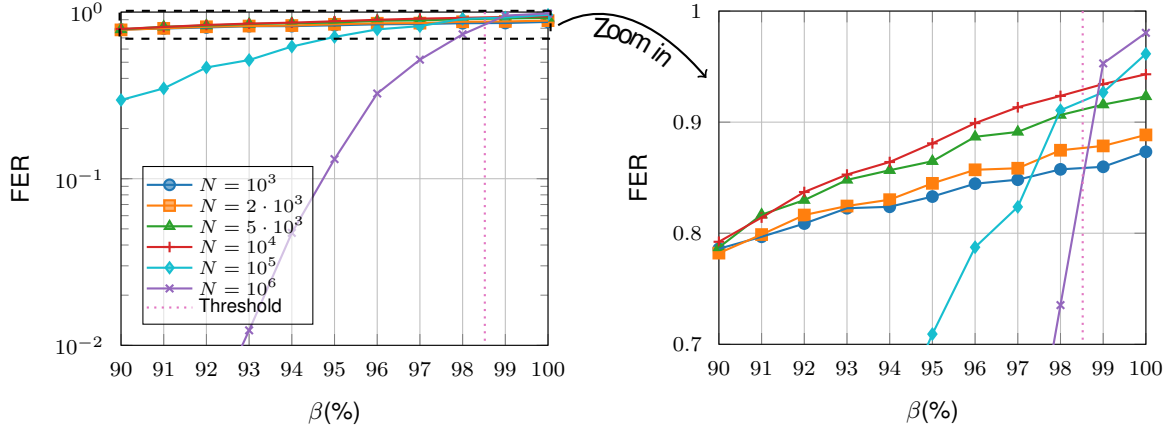
\begin{figure*}[t!]
    \centering
    \begin{tikzpicture}
\def\myshift#1{\raisebox{-1.7ex}}
    \node at (0,0) {  \begin{tikzpicture}[>=latex]
\begin{axis}[
every axis/.append style={font=\small},
tick label style={font=\footnotesize},
xlabel = $\beta$(\%),
ylabel = FER,
xmin = 90, xmax =100,
ymin = 0.01, ymax = 1,
ymode = log,
xtick distance=1,
x tick label style={yshift= -1mm},
y tick label style={xshift= -1mm},
ylabel shift = 1mm,
xlabel shift = 1mm,
set layers, mark layer=axis tick labels,
grid = major,
width = 0.45\linewidth,
 height = 6cm,
 xlabel near ticks,  
 ylabel near ticks,  
 xticklabel style={/pgf/number format/fixed},
every axis plot/.append style={thick},legend style={at={(0.05,0.3)},anchor=west, font = \scriptsize,row sep=-0.75ex,inner sep=0.2ex, fill opacity = 0.6, text opacity = 1},
legend cell align={left},
cycle list name = foo
]

\pgfplotstableread{Figures/FER.txt}
\datatable
\pgfplotsinvokeforeach {1,2,3,4,5,6}{
\addplot+
         table
         [
          x expr=\thisrowno{0}, 
          y expr=\thisrowno{#1} 
         ] {\datatable};
}
\addlegendentry{$N = 10^3$}
\addlegendentry{$N = 2\cdot10^3$}
\addlegendentry{$N = 5\cdot10^3$}
\addlegendentry{$N = 10^4$}
\addlegendentry{$N = 10^5$}
\addlegendentry{$N = 10^6$}
\addlegendimage{C7,dotted,thick}
\addlegendentry{Threshold}
\draw[dotted,thick, C7](axis cs: 98.52, 1) -- (axis cs: 98.52, 0.01);
\end{axis}
\end{tikzpicture}};
    \draw[dashed,thick] (3.45,2.69) -- (-2.22,2.69) -- (-2.22, 2.32) -- (3.45,2.32) -- (3.45,2.69);
    \draw[thick,->,postaction={decorate,decoration={text along path,text align=center,text={|\myshift|Zoom in}}}] (3.55,2.5) to [out=30,in=130] (5.5,1.8);
    \node at (8,0) {  \begin{tikzpicture}[>=latex]
\begin{axis}[
every axis/.append style={font=\small},
tick label style={font=\footnotesize},
xlabel = $\beta$(\%),
ylabel = FER,
xmin = 90, xmax =100,
ymin = 0.7, ymax = 1,
xtick distance=1,
x tick label style={yshift= -1mm},
y tick label style={xshift= -1mm},
ylabel shift = 1mm,
xlabel shift = 1mm,
set layers, mark layer=axis tick labels,
grid = major,
width = 0.45\linewidth,
 height = 6cm,
 xlabel near ticks,  
 ylabel near ticks,  
 xticklabel style={/pgf/number format/fixed},
every axis plot/.append style={thick},legend style={at={(0.5,0.23)},anchor=west, font = \scriptsize,row sep=-0.75ex,inner sep=0.2ex},
legend cell align={left},
cycle list name = foo
]

\pgfplotstableread{Figures/FER.txt}
\datatable
\pgfplotsinvokeforeach {1,2,3,4,5,6}{
\addplot+
         table
         [
          x expr=\thisrowno{0}, 
          y expr=\thisrowno{#1} 
         ] {\datatable};
}
\draw[dotted,thick, C7](axis cs: 98.52, 1) -- (axis cs: 98.52, 0.01);
\end{axis}
\end{tikzpicture}};
\end{tikzpicture}
    \caption{$\beta$ vs. FER for an $R = 
\frac{1}{50}$ TBP-LDPC code for different blocklengths $N$. The right figure zooms in on the high FER regime. The threshold of the code was determined using density evolution. \vspace{-1mm}}
    \label{fig:FER}
\end{figure*}

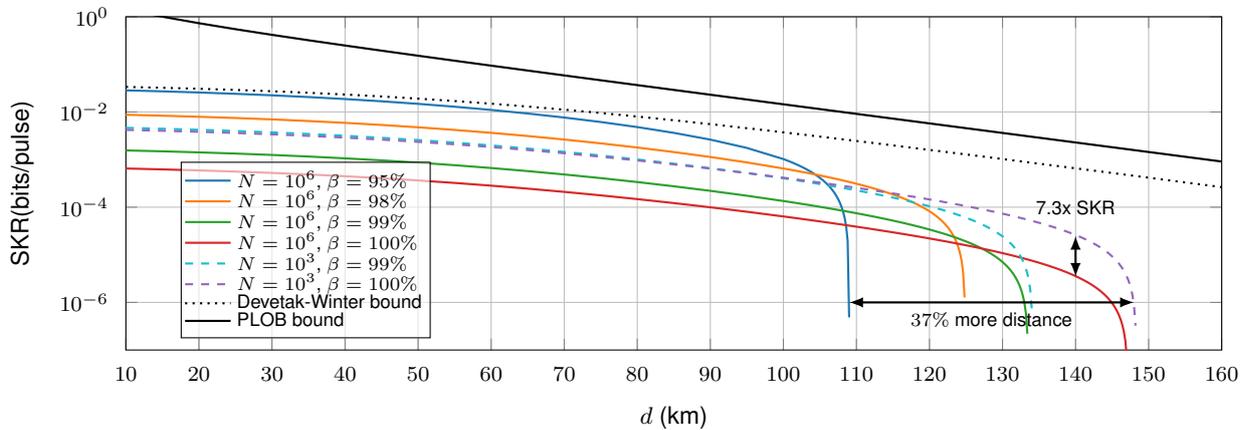
\begin{figure*}[b!]
\centering
      \begin{tikzpicture}[>=latex]
\begin{axis}[
every axis/.append style={font=\small},
tick label style={font=\footnotesize},
xlabel = $d$ (km),
ylabel = SKR(bits/pulse),
xmin = 10, xmax = 160,
ymin = 1e-7, ymax = 1,
ymode = log,
x tick label style={yshift= -1mm},
y tick label style={xshift= -1mm},
ylabel shift = 1mm,
xlabel shift = 1mm,
set layers, mark layer=axis tick labels,
grid = major,
width = \linewidth,
 height = 6cm,
 xlabel near ticks,  
 ylabel near ticks,  
 xticklabel style={/pgf/number format/fixed},
every axis plot/.append style={thick},legend style={at={(0.05,0.3)},anchor=west, font = \scriptsize,row sep=-0.75ex,inner sep=0.2ex,fill opacity = 0.6, text opacity = 1},
legend cell align={left},
cycle list name = foo
]

\pgfplotstableread{Figures/SKR_distance.txt}
\datatable
\pgfplotsinvokeforeach {4,3,2,1}{
\addplot+[no marks]
         table
         [
          x expr=\thisrowno{0}, 
          y expr=\thisrowno{#1} 
         ] {\datatable};
}

\pgfplotsinvokeforeach {6,5}{
\addplot+[no marks, dashed]
         table
         [
          x expr=\thisrowno{0}, 
          y expr=\thisrowno{#1} 
         ] {\datatable};
}

\addplot[no marks, thick, black, dotted]
         table
         [
          x expr=\thisrowno{0}, 
          y expr=\thisrowno{8} 
         ] {\datatable};
         
\addplot[no marks, thick, black]
         table
         [
          x expr=\thisrowno{0}, 
          y expr=\thisrowno{7} 
         ] {\datatable};
\addlegendentry{$N = 10^6$, $\beta = 95\%$}
\addlegendentry{$N = 10^6$, $\beta = 98\%$}
\addlegendentry{$N = 10^6$, $\beta = 99\%$}
\addlegendentry{$N = 10^6$, $\beta = 100\%$}
\addlegendentry{$N = 10^3$, $\beta = 99\%$}
\addlegendentry{$N = 10^3$, $\beta = 100\%$}
\addlegendentry{Devetak-Winter bound}
\addlegendentry{PLOB bound}
\draw[<->, thick] (axis cs: 140, 2.5e-5) -- (axis cs:140,3.45e-6);
\node at (axis cs: 140, 1e-4) {\scriptsize 7.3x SKR};

\draw[<->, thick] (axis cs: 109, 1e-6) -- node[below]{\scriptsize $37\%$ more distance} (axis cs:148,1e-6);
\end{axis}
\end{tikzpicture}
    \caption{SKR vs. $d$ comparing an $R = \frac{1}{50}$ TBP-LDPC code for $N = 10^6$ and $N = 10^3$ for different values of $\beta$. Also shown are the Devetak-Winter and PLOB bounds.}
    \label{fig:SKR}
\end{figure*}

\vspace{-2mm}
\section{Simulations}
In Fig. \ref{fig:FER}, we show the FER against the reconciliation efficiency for codes with different blocklengths. For all blocklengths, we have generated an $R = \frac{1}{50}$ TBP-LDPC code. For each point, we simulate until at least 100 frame errors have occurred. As $N$ increases, the slope of the FER curve increases, indicating a better error correction performance. For $N = 10^6$, the curve is steep and starts approaching the threshold of the code, while for smaller $N$, the FER stays high for $\beta \geq 90\%$. Note that focusing on the high FER regime, the FER of the short blocklength error correction codes is better when operating beyond the threshold of the code.

In Fig. \ref{fig:SKR} we show how this lower FER impacts the SKR over long distances for different $N$ and $\beta$. For this CV-QKD system, we use the following parameters: quantum efficiency $\eta = 0.6$, excess noise on Bob $\xi_{Bob} = 0.001$, electronic noise $\nu_{el} = 0.01$, fibre attenuation $\alpha = 0.2$ dB/km, and $N_{privacy} = 10^{10}$. The modulation variance $V_{A}$ is chosen such that at any distance $d$ the mutual information $I_{AB} = R/\beta$. For the outer code, we assume $R_{out} = 0.999$. We also show the Devetak-Winter bound \cite{devetak2005distillation}, which is a lower bound on the maximum achievable SKR, and the Pirandola, Laurenza, Ottaviani and Banchi (PLOB) bound \cite{pirandola2017fundamental}, which is the upper bound on the SKR.

As $\beta$ increases, the distance over which keys can be shared between Alice and Bob increases. The distance increase of $\beta =100\%$ compared to $\beta = 95\%$, which is a common value of $\beta$, is 37\%, however, the key rates are lower because of the high FER. When $N = 1000$, we observe an increase in SKR of 7.3 times compared to $N = 10^6$ at a distance of 140km. There is still a significant gap between the key rates reported here and the PLOB bound. A potential way to bridge the gap would be to use error correction codes with $\beta > 1$ for the first decoding step of our short blocklength reconciliation protocol, however, the security proofs are not yet mature enough to accurately estimate the key rates.

\vspace{-3mm}
\section{Conclusion}
In this paper, we have proposed a reconciliation protocol for short blocklength error correction codes ($N = 10^3$). We show that for long-distance (140km) transmission, using short blocklength error correction codes can increase SKRs by 7.3 times, while offering reduced decoding complexity. Future research will focus on designing good short blocklength error correction codes and security proofs for reconciliation with $\beta > 1$.
\newpage
\clearpage
\section{Acknowledgements}
This work was supported by the PhotonDelta GrowthFunds Programme on Photonics and by the Dutch Ministry of Economic Affairs (EZ), as part of the Quantum Delta NL KAT-2 programme on Quantum Communications. This work also acknowledges support of the European Union via the Marie Curie Doctoral Network QuNEST (Grant Agreement: 101120422) and the EIC Transition Grant project PAQAAL (under Grant Agreement 101213884)

\printbibliography

\end{document}